\documentclass[conference,onecolumn,12pt]{IEEEtran}

\usepackage[utf8]{inputenc} 
\usepackage[T1]{fontenc}
\usepackage{url}
\usepackage{ifthen}
\usepackage{cite}
\usepackage{setspace}
\usepackage[cmex10]{amsmath} 

\usepackage{amsthm}
\usepackage{mathtools}
\usepackage{amsfonts}
\usepackage{xparse}
\usepackage{physics}

\usepackage{algorithmic}

\ifdefined \theorem 
\else
  \newtheorem{theorem}{Theorem}
\fi

\newtheorem{lemma}{Lemma}
\ifdefined \corollary 
\else
\newtheorem{corollary}{Corollary}
\fi

\ifdefined \proposition 
\else
\newtheorem{proposition}{Proposition}
\fi

\ifdefined \definition 
\else
\newtheorem{definition}{Definition}
\fi

\ifdefined \problem 
\else
\newtheorem{problem}{Problem}
\fi

\ifdefined \example 
\else
\newtheorem{example}{Example}
\fi

\ifdefined \remark 
\else
\newtheorem{remark}{Remark}
\fi

\ifdefined \conjecture 
\else
\newtheorem{conjecture}{Conjecture}
\fi

\global\long\def\EE{\mathbb{E}}

\global\long\def\11{\mathbbm{1}}

\newcommand{\CM}{\mathcal{M}}

\newcommand{\CU}{\mathcal{U}}

\global\long\def\+{\oplus}

\def\<{\langle}
\def\>{\rangle}

\ifdefined \abs \else
 \newcommand{\abs}[1]{\lvert#1\rvert}
\fi

\ifdefined \norm \else
 \newcommand{\norm}[1]{\lVert#1\rVert}
\fi

\ifdefined \set 
  \renewcommand{\set}[1]{\left\{#1\right\}}
\else
  \newcommand{\set}[1]{\left\{#1\right\}}
\fi

\DeclareMathOperator*{\polylog}{polylog}

\def\zo{\{0,1\}}

\providecommand{\tr}{tr}

\ifdefined \Tr 
  \renewcommand{\Tr}[1]{\tr \Big\{#1\Big\}}
\else
  \newcommand{\Tr}[1]{\tr \Big\{#1\Big\}}
\fi

\usepackage[nolist,nohyperlinks]{acronym}

\newacro{ptp}[PtP]{Point-to-Point}
\newacro{iid}[i.i.d.]{independent and identically distributed} 
\newacro{IID}[i.i.d.]{independent and identically distributed} 
\newacro{PAC}[PAC]{\textit{probably approximately correct}}
\newacro{VC}[VC]{Vapnik–Chervonenkis}
\newacro{ERM}[ERM]{\textit{empirical risk minimization}}
\newacro{SVM}[SVM]{support-vector machine}
\newacro{POVM}[POVM]{positive operator-valued measure}
\newacro{QPAC}[QPAC]{\textit{quantum probably approximately correct}}

\newacro{QSRM}[QSRM]{\textit{quantum shadow risk minimization}}
\newacro{QC}[QC]{quantum computer}
\acrodefplural{OC}[OC's]{quantum computers}
\newacro{ML}[ML]{machine learning}
\newacro{QML}[QML]{quantum-enhanced machine learning}
\newacro{NISQ}[NISQ]{noisy intermediate-scale quantum}
\newacro{VQA}[VQA]{variational quantum algorithms}
\newacro{QNN}[QNN]{quantum neural network}
\newacro{CST}[CST]{classical shadow tomography}
\newacro{QST}[QST]{quantum state tomography}

\begin{document}
\title{Improved Classical Shadow Tomography Using Quantum Computation} 

\author{%
 \IEEEauthorblockN{Zahra Honjani and Mohsen Heidari}
\IEEEauthorblockA{Department of Computer Science, Indiana University\\
Email: \tt \{zhonjani, mheidar\}@iu.edu}
}

\maketitle
\thispagestyle{plain}

\begin{abstract}
   Classical shadow tomography (CST) involves obtaining enough classical descriptions of an unknown state via quantum measurements to predict the outcome of a set of quantum observables. CST has numerous applications, particularly in algorithms that utilize quantum data for tasks such as learning, detection, and optimization.
    This paper introduces a new CST procedure that exponentially reduces the space complexity and quadratically improves the running time of CST with single-copy measurements. 
    The approach utilizes a quantum-to-classical-to-quantum process to prepare quantum states that represent shadow snapshots, which can then be directly measured by the observables of interest. 
    With that, calculating large matrix traces is avoided, resulting in improvements in running time and space complexity. 
    The paper presents analyses of the proposed methods for CST, with Pauli measurements and Clifford circuits.
   
    
\end{abstract}

\section{Introduction}
\Ac{QST} is a procedure to obtain an approximate description of an unknown state $\rho$ through repeated measurements on several copies of the state. As a result, the expectation of any observable can be approximated by direct computation on a classical computer. In general, the copy complexity of  \ac{QST} for  
a $d$-dimensional mixed state $\rho$ scales as $\Theta(d^2)$ \cite{Haah2016}. Unfortunately, this means that the copy complexity scales exponentially with the number of qubits since $d = 2^n$. Additionally, $2^{\Omega(d)}$ classical bits are needed to store the full description of the state. 
Therefore, \ac{QST} may be unnecessary when the goal is to predict only a specific class of observables. 
Holevo, in 1973, stated that Alice can communicate at most $n$ bits of information by sending $n$ qubits to Bob \cite{holevo1973estimates}. Later, in 2018, Aaronson built upon this idea and introduced shadow tomography as a more efficient procedure to approximate the acceptance probability of 2-outcome measurements performed on an unknown state $\rho$ \cite{Aaronson2018}. Particularly, it was shown that $\tilde{O}(\epsilon^{-4} \log^4 M  \log d)$ copies of $\rho$ are sufficient to approximate the acceptance probability of any set of $M$ measurements within an additive error $\epsilon>0$ ---  hence a logarithmical reduction in the number of copies. 
Although this approach is highly sample-efficient, it imposes significant demands on quantum hardware, requiring joint measurements of multiple copies of $\rho$ stored in quantum memory.
Later, Huang et al. introduced \ac{CST}, which provides a minimal classical sketch of $\rho$ that can be used to predict several observables \cite{Huang2020}. The authors presented a protocol with single-copy measurements to construct a rough classical description of $\rho$ as a $d\times d$ matrix $\hat{\rho}$. This matrix is then used to compute the approximate expectation of given observables $O_1, ..., O_M$ by directly computing $\tr(\hat{\rho}O_j)$ followed by a median-of-means (MoM) estimator. It was shown that $O(\frac{B}{\epsilon^2}\log M )$ copies of $\rho$ is sufficient to achieve an $\epsilon$-approximation, where $B$ is the maximum {\emph{shadow norm}} of the given observables. \ac{CST} has found widespread success in various applications, including measurement of electronic Hamiltonians \cite{Hadfield2021adaptive, Hadfield2020local, Huang2021preskill}, fidelity approximation \cite{Huang2020}, quantum learning and databases \cite{Zhang2025,huang2022learning,Heidari2024,Huang2021preskill}.   While reducing the quantum hardware demand, the runtime of CST is generally demanding as it relies on the classical computation of traces of matrices with exponential dimensions.   
Additionally, space complexity, measured in bits, is another crucial resource consideration that is sometimes overlooked. The mentioned procedures require exponential bits of space to perform the shadow tomography. 
In general, the existence of a hyper-efficient shadow tomography with running time and copy complexity scaling as $\polylog(M, d)$ appears to be unrealistic as it implies $BQP/qpoly=BQP/poly$, which was shown to be false \cite{aaronson_kuperberg_2006}. Consequently, any advancements in time, space, or copy complexity must remain within these constraints. This result has shifted the research focus toward special cases of shadow tomography.
In light of this, we explore improvements of \ac{CST} in terms of running time and space complexity for special classes of observables. Our proposed solution achieves an exponential reduction in overall (classical) space complexity and a quadratic improvement in running time. We propose a quantum-to-classical-to-quantum (QCQC) approach whereby, after measuring copies of $\rho$, a classical-quantum (CQ) channel is applied to prepare a quantum state representing a shadow snapshot. This state is then measured using the given observables implemented as circuits on a quantum computer. By measuring this state directly, one avoids the exponential classical computation for approximating the trace, given that the gate complexity of the observables is polynomial. The CQ channel design depends on the choice of random measurements within the CST process. We analyze the performance of the proposed approach for two classes of choices: random Pauli measurements and Clifford measurements. The former is suitable for local observables, whereas the latter is suitable for low-rank observables.

\subsection{Other related works}
Below, we highlight some of the most relevant papers to this work. 
Different schemes have been studied for shadow tomography, for instance, \cite{Chen2024,Robbieefficientshadow} explored shadow tomography with measurements that entangle only a constant number of copies of $\rho$ at a time, while Acharya et al. employed informationally complete positive operator-valued measure for correlation functions related to quantum states \cite{Acharyashadow}. {\emph{Shallow shadows}} uses a depth-modulated randomized measurement scheme, which is a general format for Pauli and Clifford measurement schemes in \ac{CST} \cite{bertoni2024shallow}. Becker et al. obtained an approach for \ac{CST} of continuous value quantum state using randomized Gaussian unitaries and measurements, which can estimate the expectation value of special canonical observables with states with bounded moments within a polynomial-time calculation \cite{Becker2024CST}. Koh et al. have studied \ac{CST} with noisy circuits \cite{Koh2022classical}.

\section{Preliminaries and Model Formulation}
\noindent\textbf{Notations:}  For shorthand, denote $[d]$ as $\set{1,2,...,d}$. We use the soft big-O notation $\tilde{O}$ to suppress polylogarithmic factors, that is $\tilde{O}(f(n))= O(f(n)\polylog n)$. 
The identity unitary is denoted by $\mathbb{I}$. Conjugate transpose of a unitary $U$, is defined by $U^\dagger=(U^{*})^T$. 
Given $p\in (0,1)$, the depolarizing map is defined as $$D_p(A) = pA + (1-p) \frac{\tr(A)}{d} \mathbb{I},$$ for any operator $A$ on  the underlying $d$-dimensional Hilbert space.  
Pauli gates are the set of Pauli-X, Pauli-Y, and Pauli-Z. 1-qubit Clifford gates are generated by the Hadamard (H), Phase gate (S), and CNOT gate. The Clifford group on $n$ qubits consists of all unitary operations that map tensor products of Pauli operators to other tensor products of Pauli operators. 


\subsection{Problem Formulation}
The main problem of interest in this work is described as follows. Consider a set of observables $ \{ O_1, O_2, ..., O_M\}$ in a $d$-dimensional quantum system. The objective is to approximate $o_j:=\tr(O_j \rho)$ for all $j\in [M]$, which is the expected value of $O_j$ for an unknown mixed state $\rho$. 

\begin{problem}[Shadow Tomography]
\label{pb:shadow_tomography}
Consider a set of observables $ \{ O_1, O_2, ..., O_M\}$ in a $d$-dimensional quantum system and an unknown $d$-dimensional mixed state $\rho$. Let  ${o_j:=\tr(O_j \rho)}$ for all $j\in [M]$.  Given $\epsilon>0$, and $\delta\in (0,1)$, find a procedure that, with probability at least $1-\delta$, outputs a set of numbers $\hat{o}_j$ such that $\abs{ o_j- \hat{o}_j} \leq \epsilon$ for all $j \in [M]$.   
\end{problem}


In the following, we describe Huang et al. \cite{Huang2020}  procedure which builds the foundation of our work in solving the above problem.

\subsection{Classical Shadow Tomography} 
The \ac{CST} procedure of \cite{Huang2020} relies on a series of classical descriptions of $\rho$, which are $d\times d$ matrices obtained through random measurements on several copies of $\rho$. 
For that purpose, a unitary $U$ is randomly selected from a set $\CU$ of choices each time. This unitary is applied to $\rho$ and the system is measured in the computational basis, yielding a binary string $b\in \zo^n$. Then, one constructs the classical vector representing  $U^\dagger\ket{b}$ and the matrix corresponding to $U^{\dagger}\ketbra{b}U$. Lastly, the classical shadow snapshot is computed as the matrix $\hat{\rho} = \mathcal{M}^{-1}(U^{\dagger}\ketbra{b}U)$, where $\CM^{-1}$ is the inverse of the linear map defined as 
\begin{IEEEeqnarray}{c}
    \mathcal{M}(A) \coloneq \EE_{U\sim \CU} \biggr[\sum_{b \in \{0,1\}^n} \bra{b}U A U^{\dagger}\ket{b}U^{\dagger}\ketbra{b}U\biggr],
\label{eq:measur_ch}
\end{IEEEeqnarray}
for any operator $A$, where the expectation is taken over the choice of random unitaries $U$ from $\CU$. It is important to note that $\mathcal{M}^{-1}$ is a classical post-processing step done after the measurements. Repeating the above procedure for each copy of $\rho$ produces a sequence of shadow snapshots $\hat{\rho}_1, \cdots, \hat{\rho}_N$. Then, one computes the trace quantities $\tr(O_j \hat{\rho}_i)$ for each shadow and observable $O_j$. Finally, a MoM estimator is used to approximate each observable expectation value. 
The following results prove that \ac{CST} can be used to approximate the expectation value of several observables.
\begin{theorem}[\cite{Huang2020}]
    Problem \ref{pb:shadow_tomography} can be solved with  ${N  = O( \tfrac{1}{\epsilon^2} \log M  \max_j \norm{O_j}_{shadow}^2)}$ copies of $\rho$, where $\norm{O_j}_{shadow}$ is the {\emph shadow norm}  defined for the ensemble of unitary transformations $\CU$ used to create the classical shadows.
\end{theorem}

Two special sets $\CU$ for the choice of unitary transformation have been studied thoroughly. The first case is CST with Pauli measurements, which is suitable for local observables. Here,  $\CU$ is the set of all tensor products of 1-qubit Clifford gates that is equivalent to measuring each qubit on a random Pauli basis.  Let the chosen unitary be  $U = U_1 \otimes U_2 \otimes ... \otimes U_n$, where each $U_l$ is a random 1-qubit Clifford gate. The classical shadow snapshot has the following closed-form expression
    \begin{IEEEeqnarray}{c}
        \label{eq:rho_h_pauli}
        \hat{\rho} = \overset{n}{\underset{l=1}{\bigotimes}}(3U_l^{\dagger}\ketbra{b_l} U_l - \mathbb{I}).
    \end{IEEEeqnarray}
   Moreover, the shadow norm is bounded as
   \begin{IEEEeqnarray}{c}
       \norm{O_j}_{shadow} \leq 2^{k} \|O_j\|_\infty,
       \label{eq:pa_norm}
   \end{IEEEeqnarray}
    where $k$ is the locality of the observable. As a result, if all observables have locality at most $k$, then the copy complexity of CST with Pauli measurements scales as ${N= O(\tfrac{1}{\epsilon^2} 4^k \log M \max_j \|O_j\|_{\infty}^2)}$. As for the running time, for each $j\in [M]$ and $i\in [N]$, computing $\tr(O_j \hat{\rho}_i)$ takes $O(2^{2k})$ time. Moreover, the MoM for $K$ groups takes $O(K \log K)$. Thus, the total runtime of CST with Pauli measurements for bounded observables is $O(16^{k} M \log M).$
   

Another choice of unitaries is the class of Clifford $n$-qubit circuits. In this case, the classical shadow snapshot is given by
    \begin{IEEEeqnarray}{c}
        \hat{\rho} = (2^n+1) U^{\dagger}\ketbra{b} U - \mathbb{I},
        \label{eq:rho_cli}
    \end{IEEEeqnarray}
    and the shadow norm of observable $O_j$ is bounded as 
    \begin{IEEEeqnarray}{c}
        \norm{O_j}_{shadow} \leq \sqrt{3\tr(O_j^2)}.
        \label{eq:shadow_clif}
    \end{IEEEeqnarray}
    This means that CST with Clifford measurements is useful for non-local observables with bounded Hilbert-Schmidt norm.   Low-rank bounded observables are special cases of such operators.   An example is the fidelity measuring problem where $O_j=\ketbra{\psi_j}$ for some states $\ket{\psi_j}$. Hence, $\tr(O_j^2)=1$ and the sample complexity has a nice form $O(\tfrac{1}{\epsilon^2} \log M)$ which scales logarithmically with $M$. However, the time complexity of CST in this case is $\Omega(d^2 M \log M)$, which is exponential in the number of qubits. 

In this work, we study improvements in running time and space complexity of CST through a QCQC approach, in which parts of the classical computations are done on a quantum computer, resulting in a computational speedup.

 \section{Main Results}
We first propose our QCQC approach for CST with Clifford measurements and present our theoretical analyses. In Section \ref{subsec:CST pauli}, we discuss our approach for CST with Pauli measurements. 

Our  QCQC approach to CST  consists of two phases.
The first phase is the same as CST procedure: a unitary $U$ is randomly selected from a set $\CU$, applied to the state $\rho$, followed by a measurement in the computational basis. The binary outcome $b\in \zo^n$, and description of $U$, are stored classically.  
In the second phase,  deviating from the original CST, the measurement outcomes are used to prepare a quantum state. Then, a series of quantum circuits are applied, followed by measurements with the given observables $O_j$ implemented in a quantum computer. The measurement results of the second phase are then post-processed to estimate $o_j$. 

 We show that the second phase effectively reduces the computational time and space required for CST without increasing the sample complexity. We propose different methods for the second phase of the procedure. Below, we investigate the QCQC approach with n-qubit Clifford (Clifford measurements) and single-qubit Clifford (Pauli measurements) circuits as two choices of $\CU$.

\subsection{QCQC with Clifford Measurements}
As described above, in the first phase, a random n-qubit Clifford gate $U_i$ is applied to the $i$'th copy of the input state. This is followed by measurement in the computational basis, resulting in $b_i\in \zo^n$. For the second phase, the state $\ket{b_i}$ is prepared, and the circuit for $U_i^{\dagger}$ is applied to the state, followed by measurement with observable $O_j$. Lets $X_{i,j}$ denote the resulted empirical average of $O_j$ with $m=2(d+1)$ measurement shots. It is straightforward to see that 
\begin{IEEEeqnarray}{c}\label{eq:Exp X}
    \EE[X_{i,j}] =  \tr( \EE\qty[U_{i}^{\dagger} \ketbra{b_{i}} U_{i}] O_j),
\end{IEEEeqnarray}
where the expectation is taken with respect to the randomness of generating $U_i$ and $b_i$ during the first phase. Let ${Y_{i,j} = (d+1) X_{i,j} - \tr(O_j)}.$ This process is repeated for all $i\in [N]$ and $j\in [M]$. Lastly, for each $j$ the MoM is applied on all 
$Y_{i, j}, i\in [N]$ to estimate $o_j$. In what follows, we analyze this estimation. 

First, we show that the estimator is unbiased. 
\begin{lemma} \label{lemma:Y_clif}
    $\EE[Y_{i,j}] = o_j = \tr(\rho O_j),$ $\forall i\in [N], j\in [M]$.
\end{lemma}
    \begin{IEEEproof}
    From the definition of $Y_{i,j}$ and \eqref{eq:Exp X}, we have that 
            \begin{align*}
            \EE[Y_{i,j}] &=  \EE[(d + 1)X_{i,j} - \tr(O_j)] \\
            &= \EE\qty[(d + 1)\tr( U_i^{\dagger} \ketbra{b_i} U_i O_j) - \tr(O_j)] \\
            & = \EE\qty[\tr( ((d + 1) U_i^{\dagger} \ketbra{b_i} U_i - \mathbb{I} )O_j)].
        \end{align*}
From the definition of CST channel $\CM$ for Clifford measurements in \eqref{eq:rho_cli}, the RHS of the previous equation equals $\EE[\tr( \CM^{-1}(U_i^{\dagger} \ketbra{b_1} U_i)O_j)]$. Therefore,  from the linearity of expectation and trace and the fact that $\CM^{-1}$ is a linear map, we have 
    \begin{align*}
        \EE[Y_{i,j}] & = \tr\Big(\sum_{b_i} \EE_{U_i}\bra{b_i}U_i\rho U_i^{\dagger}\ket{b_i}\mathcal{M}^{-1}(U_i^{\dagger}\ketbra{b_i}U_i)O_j\Big) \\
            &= \tr\Big(\mathcal{M}^{-1}\Big[ \sum_{b_i} \EE_{U_i}  \bra{b_i}U_i\rho U_i^{\dagger}\ket{b_i}U_i^{\dagger}\ketbra{b_i}U_i \Big] O_j\Big) \\            
            &= \tr( \mathcal{M}^{-1}[\mathcal{M}(\rho)]O_j)\\
            &= \tr(\rho O_j) = o_j,
\end{align*}
where we used the definition of $\CM$ as in \eqref{eq:measur_ch} and the distribution of $U_i$ and $b_i$.
    \end{IEEEproof}

The next lemma provides a bound for the variance of the estimation. 
\begin{lemma}
$Var[Y_{i, j}]\le 4\tr(O_j^2) $ $\forall i\in [N], j\in [M].$
    \begin{IEEEproof}
    By definition, $Var[Y_{i, j}] = (d+1)^2 Var[X_{i, j}]$.  Therefore, from the law of total variance
    \begin{align*}
        Var[X_{i,j}] = \EE[Var[X_{i,j} | U_i,b_i]] + Var[\EE[X_{i,j} | U_i,b_i]].
    \end{align*}
    The first term represents the contribution of the $m$-shot measurements for a fixed $(U_i,b_i)$. The second term represents the contribution due to the randomness of $(U_i, b_i)$. 
    Note that $X_{i,j}$ is the $m$-shot empirical average of measuring with $O_j$, that is $X_{i,j} = \tfrac{1}{m}\sum_l X_{i,j,l},$ where $X_{i,j,l}$ is the measurement outcome at the $l$th shot. Therefore, conditioned on $(U_i,b_i)$,  $Var[X_{i,j} | U_i,b_i]=\tfrac{1}{m}Var[X_{i,j,l} | U_i,b_i]$ which is bounded from above by $\frac{1}{m}\EE[X_{i, j, l}^2 |U_i,b_i]$.
Therefore, 
          \begin{IEEEeqnarray*}{Cl}
           \EE[Var[X_{i,j} | U_i,b_i]]&\leq   \frac{1}{m}\EE_{(U_i, b_i)}[\EE[X_{i, j, l}^2 |U_i,b_i]]\\
           & = \frac{1}{m}\EE_{(U_i, b_i)}\Big[\tr( U_i^{\dagger} \ketbra{b_i} U_i O_j^2)\Big]\\
            &= \frac{1}{m}\tr( \EE[U_i^{\dagger} \ketbra{b_i} U_i] O_j^2) \\
            &\stackrel{(a)}{=} \frac{1}{m} \tr\Big(D_{\tfrac{1}{d+1}}(\rho)O_j^2\Big) \\
            &= \frac{1}{m (d+1)} \tr((\rho+\mathbb{I}) O_j^2),
        \end{IEEEeqnarray*}
        where (a) follows as the CST channel for Clifford measurements is the depolarizing channel $D_{\tfrac{1}{d+1}}$ \cite[(S41)]{Huang2020}.  Moreover, because of the identity  $\norm{A}=\max_\sigma \tr(A\sigma)$, the trace quantity in the RHS is not greater than $\norm{O_j^2}+\tr(O_j^2)\leq 2 \tr(O_j^2)$.    
As for the second term, $\EE[X_{i,j} | U_i,b_i] = \tr( U_i^{\dagger} \ketbra{b_i} U_i O_j)$. Hence, the variance of this term is bounded by the shadow norm $\tfrac{1}{(d+1)^2}\norm{O_j}^2_{shadow}$ \cite[S7]{Huang2020}. Therefore, $Var[Y_{i, j}] \leq \frac{2(d+1)}{m} \tr(O_j^2)+ \norm{O_j}^2_{shadow}.$ Lastly, the lemma follows from \eqref{eq:shadow_clif}, and by setting $m=2(d+1)$.
    \end{IEEEproof}
    \label{lemma:clif_var}
\end{lemma}
With these two lemmas, the first main result of the paper is presented below. 
\begin{theorem}  \label{th:clif_main}
    Given  a set of observables $ \set{ O_1, O_2, ..., O_M}$ satisfying $\max_j\tr(O_j^2) = B$, and each implemented with at most $F$ quantum gates,  there is a procedure that 
    solves Problem 1 with $O(B\log M  )$ copies of $\rho$, and  $\tilde{O}(d B F M)$ quantum and classical time, and $O(\polylog(d,M)+M\log M)$  classical space. 
\end{theorem}
    \begin{IEEEproof}
        Similar to the steps used in \cite{Huang2020}, for each ${j\in[M]}$,  we use a MoM on $Y_{i,j}, {i\in [N]}$.  Particularly,  we group $Y_{i,j}, {i\in [N]}$ into $K$ groups and take the average of each group. Then, the median of the averages of each group is declared as the estimation $\hat{o}_j$ of $o_j$. From the analysis of  \cite{MOMnotes}, the estimation error satisfies
    \begin{IEEEeqnarray*}{c}
        P(\abs{\hat{o}_{j} - o_j} > \epsilon) \leq \exp{-2K(\frac{1}{2}- \frac{K}{N} \frac{Var(Y_{1,j})}{\epsilon^2})}.
    \end{IEEEeqnarray*}
     The RHS is less than equal to $\tfrac{\delta}{M}$, if  $$K = 2 \log  \frac{M}{\delta},$$ and $$N = \tfrac{8}{\epsilon^2}\log( M/\delta)  Var(Y_{1,j}).$$ A union bound over $j\in [M]$ proves that the conditions of Problem 1 are satisfied for the given $\epsilon, \delta$.  From Lemma \ref{lemma:clif_var}, $Var(Y_{1,j}) \leq 4\tr(O_j^2).$ Thus, the sample complexity of shadow tomography is obtained. As for the runtime, there exists an algorithm that samples uniformly from the Clifford group in classical time $O(n^8)$ and outputs a circuit representing the measurement with a gate complexity of $O(n^2)$, where $n = \log d$   \cite{DiVincenzo2002}. For each $i$, a Clifford circuit is constructed, and the state is measured. Then, the second phase of QCQC is repeated $M$ times, once for each observable. Each repetition takes $m F \polylog(d)$ time, with $m=O(d)$.   The running time of MoM for each $j$ is $O(K\log K)$. Hence, the overall time is $O(mNMF\polylog(d) + MK\log K)$. For each $i$, we need to store $b_i\in \zo^n$ and the description of $U_j$, which requires $\polylog(d)$ space. Moreover, $M K$ registers are needed to store the average of each group for the MoM. 
   
    \end{IEEEproof}

Compared to CST of \cite{Huang2020}, a quadratic speed up w.r.t $d$ is obtained. Because of the matrix multiplication, CST has a running time scale with $\Omega(d^2 M \log M)$. Moreover, the working space is reduced exponentially as CST requires storing the $d\times d$ matrix of each shadow snapshot $\hat{\rho}$,
whereas here, just $O(\polylog(d))$ is needed to store the unitary matrix and output of the first circuit. 

The next result shows that QCQC applies to a broad class of small-ranked observables. 
\begin{corollary}
    If the rank of each $O_j$ is bounded by $r$, and the maximum absolute value of it's eigenvalues is not greater than $\lambda$, then QCQC solves Problem 1 in $O(dr\lambda^2 M \log M)$ time with $O(r \lambda^2 \log M)$ copies. 
\end{corollary}
\subsection{QCQC with Pauli Measurements} \label{subsec:CST pauli}
Next, we explore QCQC with Pauli measurements. In the same way as CST, tensor products of 1-qubit Clifford gates are used as $U_i, i\in [N]$ in the first phase. Let $b_i\in \zo^n, i\in [N]$ be the measurement outputs in the first phase.
In the second phase, we explain how to prepare the state based on the  $b_i$ values. Each bit of  $b_i$ is flipped with a probability of $\tfrac{1}{3}$ , resulting in the noisy string $c_i := b_i \oplus e_i$, where $e_i=(e_{i,1}, ..., e_{i,n})$ and
\begin{IEEEeqnarray}{c}
    e_{i,l} =
\begin{cases}
0 & \text{with probability } \frac{2}{3}, \\
1 & \text{with probability } \frac{1}{3}.
\end{cases}
\end{IEEEeqnarray}
This randomization helps to implement the shadow snapshots as quantum states. Note that the original shadows in CST are not valid density operators as they are not positive semi-definite. 

Next, we prepare the state $\ket{c_i}$ and apply $U_i^{\dagger}$ on it, followed by measurement using each observable $O_j$. The output is $X_{i,j}$ with average
\begin{IEEEeqnarray}{c}
    \EE[X_{i,j}] = \tr(\EE[U_{i}^{\dagger} \ketbra{c_{i}} U_{i}] O_j).
\end{IEEEeqnarray}
Then, we define a weight variable as $  w_{i,l} = 3 (-1)^{e_{i,l}}$ and compute $Z_{i,j} = X_{i,j} \times \underset{l}{\prod} w_{i,l}$. For any fixed $(U_i, b_i)$ we repeat this process $m$ times, each time with $e_{i,l}$'s generated randomly and independently. Then we take the empirical average  denoted by $Y_{i,j}=\tfrac{1}{m}\sum_{r} Z_{i, j, r}$. 
Then $o_j$ will be predicted by applying the MoM on $Y_{i,j}, {i\in [N]}$. 

Now, we will prove some of the properties of this method. 
\begin{lemma} $ \EE[Y_{i, j}] = o_j$
\label{lemma:y_exp}
\end{lemma}
\begin{IEEEproof}
Set $U = \overset{n}{\underset{l=1}{\bigotimes}} U_l$ where $U_1, U_2, ..., U_n$ are randomly selected 1-qubit Clifford gates, and $b$ the output of applying $U$ on the sample $\rho$, and $O_j$ the j'th measurement operator. Then,
        \begin{IEEEeqnarray*}{Cl}
            \EE[Y_{i, j}&|b_i, U_i] = \EE[Z_{i,j} |b_i, U_i] = \EE[X_{i,j} \times \underset{l}{\prod} w_{i,l}|b_i, U_i] \\
            &=  \tr(\EE_{e_{i,l}}\qty\Big[\underset{l}{\prod} w_{i,l} \times(\overset{n}{\underset{l}{\bigotimes}} U_{i,l}^{\dagger} \ketbra{c_{i,l}} U_{i,l})]O_j)\\
            &= \tr(\EE_{e_{i,l}}\qty\Big[\overset{n}{\underset{l}{\bigotimes}} w_{i,l}(U_{i,l}^{\dagger} \ketbra{c_{i,l}} U_{i,l})]O_j)\\
            &= \tr(\overset{n}{\underset{l}{\bigotimes}}  (U_{i,l}^{\dagger} \EE_{e_{i,l}}\qty[w_{i,l} \ketbra{c_{i,l}}] U_{i,l})O_j),
        \end{IEEEeqnarray*}
where we exchanged the product with the tensor and used the fact that $e_{i,l}$ are independent for different $l$'s. Let $\Bar{b}_{i,l}$ be the complement of ${b}_{i,l}$. Then, from the definition of $e_{i,l}$ the RHS equals
          \begin{IEEEeqnarray*}{Cl}
            & \tr(\overset{n}{\underset{l}{\bigotimes}}  (U_{i,l}^{\dagger} (2 \ketbra{b_{i,l}} -  \ketbra{\Bar{b}_{i,l}}) U_{i,l})O_j)\\
            &=  \tr(\overset{n}{\underset{l}{\bigotimes}}  (3U_{i,l}^{\dagger}\ketbra{b_{i,l}} U_{i,l} - \mathbb{I})O_j),
        \end{IEEEeqnarray*}
        where we used the identity $\mathbb{I} = \ketbra{b_{i,l}} + \ketbra{\Bar{b}_{i,l}}$.
        Using the definition of CST channel for Pauli measurements, the RHS of the previous equation equals 
        $ \tr(\mathcal{M}^{-1}(U^{\dagger}\ketbra{b}U) O_j)$. Therefore,  taking the expectation over $U, b$ gives
        \begin{IEEEeqnarray*}{Cl}
            \EE[Y_{i,j}] &= \EE_{U,b}[\EE[Y_{i,j}|b, U]] \\
            &=  \tr(\EE_{U}\underset{b}{\sum} \bra{b}U\rho U^{\dagger}\ket{b}\mathcal{M}^{-1}(U^{\dagger}\ketbra{b}U)O_j) \\
            &=  \tr( \mathcal{M}^{-1}\biggr[ \EE_{U}\underset{b}{\sum}  \bra{b}U\rho U^{\dagger}\ket{b}U^{\dagger}\ketbra{b}U \biggr] O_j) \\
            &=  \tr( \mathcal{M}^{-1}[\mathcal{M}(\rho)]O_j) \\
            &=  \tr( \rho O_j) =  o_j.
        \end{IEEEeqnarray*}
    \end{IEEEproof}
    
\begin{lemma}
$Var[Y_{i,j}]\le \frac{9^n}{m}\norm{O_j}^2 + \norm{O_j}^2_{shadow}.$
    \label{lemma:var_pauli}
\end{lemma}
    \begin{IEEEproof}
        Using law of total variance, $Var[Y_{i,j}] = \EE[Var[Y_{i,j} | U_i,b_i]] + Var[\EE[Y_{i,j} | U_i,b_i]]$. As for the first term, conditioned on $(U_i,b_i),$ and  the definition of $c_{i,l}$, 
        \begin{IEEEeqnarray*}{Cl}
             \EE[&Y_{i,j}^2| U_i,b_i] =\frac{1}{m} \EE_{e_{i,l}} [X_{i,j}^2 \times \prod_l w_{i,l}^2| U_i,b_i]\\
             & =\frac{9^n}{m} \EE_{e_{i,l}} [X_{i,j}^2 | U_i,b_i] \\
             &=\frac{9^n}{m} \tr\Big( \EE_{e_{i,l}} \qty\Big[\bigotimes_l (U_{i,l} ^\dagger \ketbra{c_{i,l}}  U_{i,l})] O_j^2\Big) \\
             &=\frac{9^n}{m} \tr\Big(\bigotimes_{l} (U^\dagger (\frac{2}{3} \ketbra{b_{i,l}}+\frac{1}{3} \ketbra{\Bar{b}_{i,l}}) U_l) O_j^2\Big) \\
             &= \frac{9^n}{m}\tr\Big(\bigotimes_{i,l} (\frac{1}{3} U_{i,l}^\dagger \ketbra{b_{i,l}} U_{i,l} +  \frac{1}{3}\mathbb{I}) O_j^2\Big).
        \end{IEEEeqnarray*}
Note that the trace quantity in the RHS equals $\tr(D_{1/3} ^{\otimes n} (U^\dagger \ketbra{b} U) O_j^2)$. Therefore, from the inequality $Var(T)\leq \EE[T^2]$, the first variance component of $Y_{i,j}$ is bounded as 
        \begin{IEEEeqnarray*}{Cl}
            \EE[Var[Y_{i,j} | U_i,b_i]] & \leq  \frac{9^n}{m}\EE_{U,b} \qty[\tr(D_{1/3} ^{\otimes n} (U^\dagger \ketbra{b} U) O_j^2)] \\
            &= \frac{9^n}{m}\tr(D_{1/3} ^{\otimes n} (\EE_{U,b}[U^\dagger \ketbra{b} U]) O_j^2) \\
            &= \frac{9^n}{m}\tr(D_{1/3} ^{\otimes n} (\CM(\rho)) O_j^2) \\
            &= \frac{9^n}{m}\tr(D_{1/3} ^{\otimes n} (D_{1/3} ^{\otimes n}(\rho)) O_j^2).
        \end{IEEEeqnarray*}
        Hence, the RHS equals to $\frac{9^n}{m}\tr(D_{1/9} ^{\otimes n} (\rho) O_j^2)\leq \frac{9^n}{m}\norm{O_j}^2.$

        As for the second component of the variance, from the proof of Lemma \ref{lemma:var_pauli},  $\EE[Y_{i,j} | U_i,b_i] = \tr(\mathcal{M}^{-1}(U^{\dagger}\ketbra{b}U) O_j)$. Hence, the variance of this term can be upper bounded by the shadow norm per \cite[Lemma 1]{Huang2020}.

        As a result, we have shown that $Var(Y_{i,j})\leq \frac{9^n}{m}\norm{O_j}^2 + \norm{O_j}^2_{shadow}.$
    \end{IEEEproof}
We build upon the previous lemmas and prove our results for observables with locality $k$, that is, observables that act non-trivially on $k$ out of $n$ qubits. 
\begin{theorem}
    Suppose $ \{ O_1, O_2, ..., O_M\}$ have locality at most $k$ (a.k.a $k$-junta) and $B = \max_j \norm\big{O_j}^2$. There is a procedure that 
    solves Problem 1 with  $O(9^k kB F M\log M)$ quantum and classical time, where $F$ is the maximum gate complexity of $O_j, j\in [M]$. Moreover, $O(k + M\log M)$ (working) classical space and $N = O(4^k \log M  B)$ copies of $\rho$ are sufficient for this task.      \label{th:main}
\end{theorem}
\begin{IEEEproof}
    We apply QCQC only on the relevant $k$ qubits. Suppose $O_j = O^Q_j \otimes \mathbb{I}_{Q^c}$ where $O^Q$ acts only on $Q = \{q_1, q_2, ..., q_k \}$ and $\mathbb{I}$ is the identity matrix. Let $Y^Q_{i,j}$ denote the outcome of QCQC only on the registers in $Q$. We apply MoM on these variables to generate the prediction of $o_j$.  The following shows that $ \EE[Y^{Q}_{i,j}] = o_j$.  From the proof steps of  Lemma \ref{lemma:y_exp}, it is not difficult to see that 
        \begin{IEEEeqnarray*}{Cl}
            \EE[Y^{Q}_{i,j}] &= \EE[X_{Q} \times \underset{k \in Q}{\prod} w_{k}] \\
            &= \EE_{U,b}\qty[\tr(\underset{k \in Q}{\bigotimes}  (3U_{i,k}^{\dagger}\ketbra{b_{i,k}} U_{i,k} - \mathbb{I}) O^{Q}_{i,j})] \\
            & = \EE_{U,b}\qty[\tr(\overset{n}{\underset{l=1}{\bigotimes}}  (3U_{i,l}^{\dagger}\ketbra{b_{i,l}} U_{i,l} - \mathbb{I}) O^{Q}_{j} \otimes \mathbb{I}_{Q^c})],
        \end{IEEEeqnarray*}
        where we used the fact that $\tr(3U_{i,l}^{\dagger}\ketbra{b_{i,l}} U_{i,l} - \mathbb{I})=1$. 
    From the definition of the classical shadow snapshots, the RHS of the previous equation equals ${\EE_{U,b}[\tr(\hat{\rho}O_j)]= \tr(\rho O_j) = o_j}$.
    From Lemma \ref{lemma:var_pauli}, for locality $k$,  $$Var[Y^Q_{i,j}]\le \frac{9^k}{m}\norm{O^Q_j}^2 + \norm{O^Q_j}^2_{shadow}.$$ From \eqref{eq:pa_norm}, and by setting $m=(\tfrac{9}{4})^k$, we obtain that $Var[Y^Q_{i,j}]\leq 2\times 4^k B$.  This bound is used to analyze the MoM. Similar to the analysis of MoM in Theorem \ref{th:clif_main}, it is not difficult to see that $N=O(4^k \log M  B)$, ignoring $\epsilon, \delta$.  Each round of QCQC takes $mkF$ to obtain each $Y^Q_{i,j}$. This is repeated $NM$ times, followed by the MoM. Therefore, the overall running time is $O(mNMF k + MK\log K) = \tilde{O}(9^k  F k B M)$. Moreover, for each sample, we need  $O(k)$ space to store $b_{i}$ and $U_{i}$. We store $K$ groups of empirical averages for each shadow snapshot for each observable, hence needing  $O(MK)$ classical space. Given that $K=O(\log M)$, the total space needed is $O(k M\log M)$.
\end{IEEEproof}

\section{Conclusion}
A novel quantum-to-classical-to-quantum approach is proposed for the shadow tomography problem.
The performance of this method is thoroughly evaluated in terms of running time, space complexity, and copy complexity for specific classes of observables, including local and bounded norm operators.
It is shown that this method provides a quadratic speedup in running time and an exponential improvement in classical space complexity compared to existing works.
\section*{Acknowledgment}
 This work was partially supported by the NSF Grant CCF-2211423.
\bibliographystyle{IEEEtran}

\end{document}